\documentclass[manuscript]{emulateapj}

\usepackage{amssymb}
\usepackage[export]{adjustbox}
\usepackage{amsmath}
\usepackage{booktabs}
\usepackage{hyperref}
\usepackage{multirow}
\usepackage{times}
\PassOptionsToPackage{hyphens}{url}\usepackage{hyperref}
\usepackage[hyphenbreaks]{breakurl}

\newcommand{\xmm}{{\it XMM-Newton}}

\newcommand{\arcs}{\hbox{$^{\prime\prime}$}}

\newcommand{\ls}
{\mathrel{\hbox{\rlap{\hbox{\lower4pt\hbox{$\sim$}}}\hbox{$<$}}}}
\newcommand{\gs}
{\mathrel{\hbox{\rlap{\hbox{\lower4pt\hbox{$\sim$}}}\hbox{$>$}}}}

\received{}
\begin{document}
\title{A new relativistic component of the accretion disk wind in PDS 456.}
\shorttitle{A relativistic wind in PDS 456.}
\shortauthors{Reeves et al.}
\author{J. N. Reeves\altaffilmark{1}, V. Braito\altaffilmark{1,2}, E. Nardini\altaffilmark{3}, 
A. P. Lobban\altaffilmark{4}, G. A. Matzeu\altaffilmark{5}, M. T. Costa\altaffilmark{4}}
\altaffiltext{1}{Center for Space Science and Technology, 
University of Maryland Baltimore County, 1000 Hilltop Circle, Baltimore, MD 21250, USA; email jreeves@umbc.edu}
\altaffiltext{2}{INAF, Osservatorio Astronomico di Brera, Via Bianchi 46 I-23807 Merate (LC), Italy}
\altaffiltext{3}{INAF - - Osservatorio Astrofisico di Arcetri, Largo Enrico Fermi 5, I-50125 Firenze, Italy}
\altaffiltext{4}{Astrophysics Group, School of Physical and Geographical Sciences, Keele 
University, Keele, Staffordshire, ST5 5BG, UK}
\altaffiltext{5}{European Space Astronomy Centre (ESA/ESAC), E-28691 Villanueva de la Canada, Madrid, Spain}

\begin{abstract}

Past X-ray observations of the nearby luminous quasar PDS 456 (at $z=0.184$) have revealed a wide angle accretion disk wind \citep{Nardini15}, with an outflow velocity of $\sim-0.25c$. Here we unveil a new, relativistic component of the wind through hard X-ray observations with {\it NuSTAR} and {\it XMM-Newton}, obtained in March 2017 when the quasar was in a low flux state.
This very fast wind component, with an outflow velocity of $-0.46\pm0.02c$, is detected in the 
iron K band, in addition to the $-0.25c$ wind zone. 
The relativistic component may arise from the innermost disk wind, launched from close to the black hole at radius of $\sim10$ gravitational radii.
The opacity of the fast wind also increases during a possible obscuration event lasting for 50\,ks. We suggest that the very fast wind may only be apparent during the lowest X-ray flux states 
of PDS\,456, becoming overly ionized as the luminosity increases. Overall, the total wind power may even 
approach the Eddington value.
\end{abstract}










\section{Introduction}

Outflows are an important phenomenon in Active Galactic Nuclei (AGN) and can play a key role in the
co-evolution of the massive black hole and the host galaxy \citep{DiMatteo05,King10}.
Black holes
grow by accretion and strong nuclear outflows can 
quench this process by shutting off their supply of matter, thereby setting the 
$M-\sigma$ relation that is seen today \citep{FerrareseMerritt00,Gebhardt00}.
A number of high column density ($N_{\rm H}\sim 10^{23}$\,cm$^{-2}$), fast ($\sim0.1c$)
outflows have now been found in luminous AGN \citep{Tombesi10,Gofford13}, 
through detections of blue-shifted Fe K absorption, at rest-frame energies greater than 7 keV. These ultra fast outflows may be the missing link in the 
galactic feedback process, by driving massive molecular outflows out to large ($\sim$\,kpc) 
scales in galaxies \citep{Tombesi15,Feruglio15}.

A prototype ultra fast outflow occurs in the nearby ($z=0.184$) quasar, PDS\,456.
With a bolometric luminosity of $\sim10^{47}$\,erg\,s$^{-1}$, PDS\,456 is the most luminous QSO in the local Universe \citep{Torres97, Simpson99, Reeves00} and the radio-quiet analogue of 3C\,273. However, PDS\,456 is most notable for its powerful and fast ($\sim0.25c$) X-ray wind. 
Indeed, since its initial detection in 2001 with \xmm\ 
\citep{Reeves03}, the presence of the ultra fast outflow in PDS\,456 has now been established through over a decade's worth of X-ray observations \citep{Reeves09,Behar10,Gofford14,Nardini15,Hagino15,Matzeu16,Matzeu17,Parker18}. 
Intriguingly, \citet{Hamann18} recently claimed a fast UV counterpart to the X-ray wind.

In a series of five {\it XMM-Newton} 
and {\it NuSTAR} observations of PDS\,456 in 2013--2014, \citet{Nardini15}
detected a persistent P-Cygni-like profile from highly ionized (He or H-like) iron, blueshifted to 9\,keV in the quasar rest frame. The broad P-Cygni profile established the wide-angle character of the outflow, while the wind variability 
provided a robust estimate of the wind radial distance, on the scale of the AGN accretion disk.
From this, the large mass outflow rate inferred, of $\sim10 M_{\odot} $\,yr$^{-1}$, implied that the wind power is at least 15\% of the Eddington luminosity. This is more than sufficient to provide the feedback required by models of black hole and host galaxy co-evolution \citep{HopkinsElvis10}, which likely plays a critical role in black hole growth and feedback in the early Universe. 
 

\section{Observations and Data Reduction}

\begin{figure}
\begin{center}
\rotatebox{270}{\includegraphics[height=8.5cm]{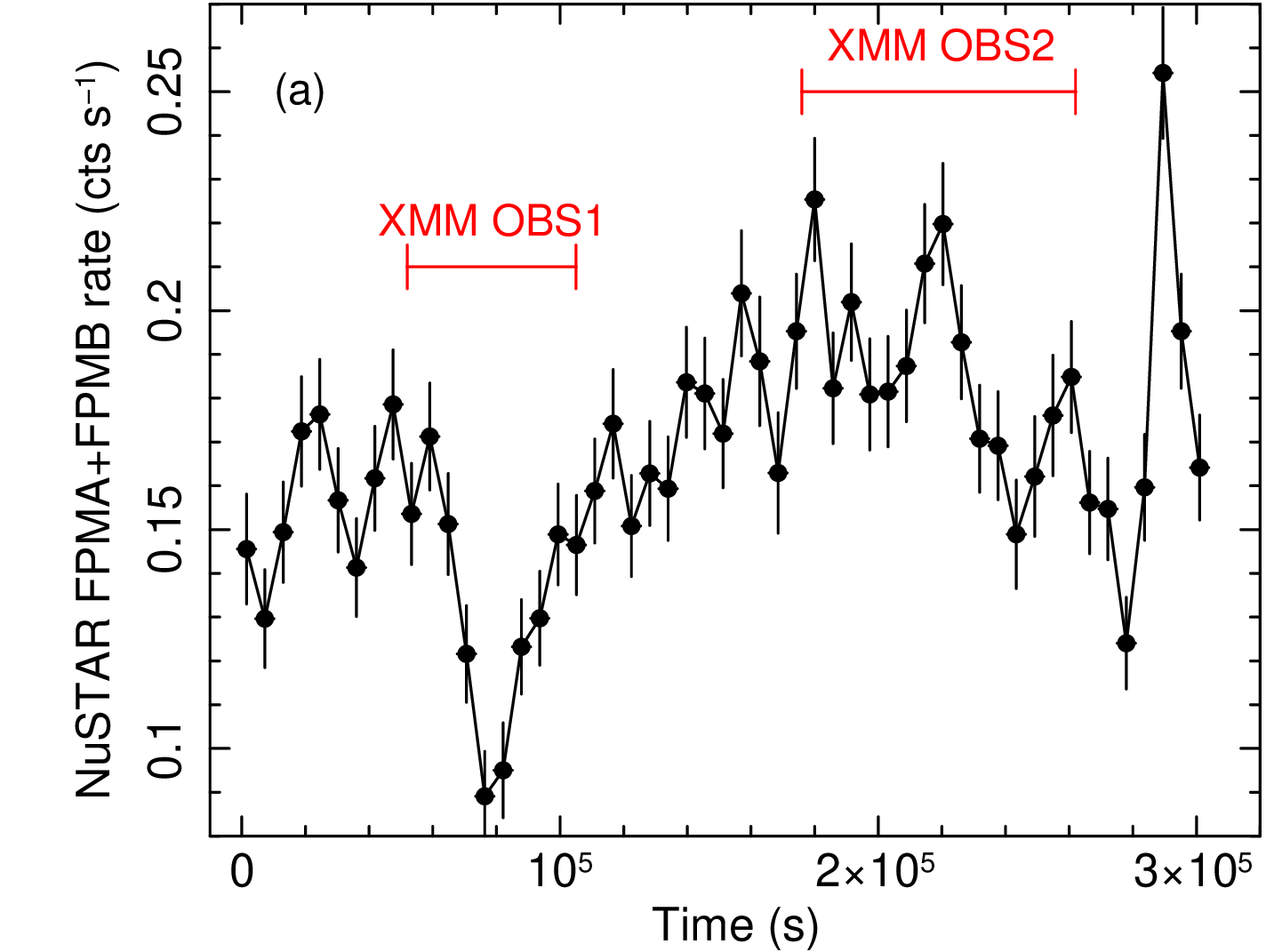}}
\rotatebox{270}{\includegraphics[height=8.5cm]{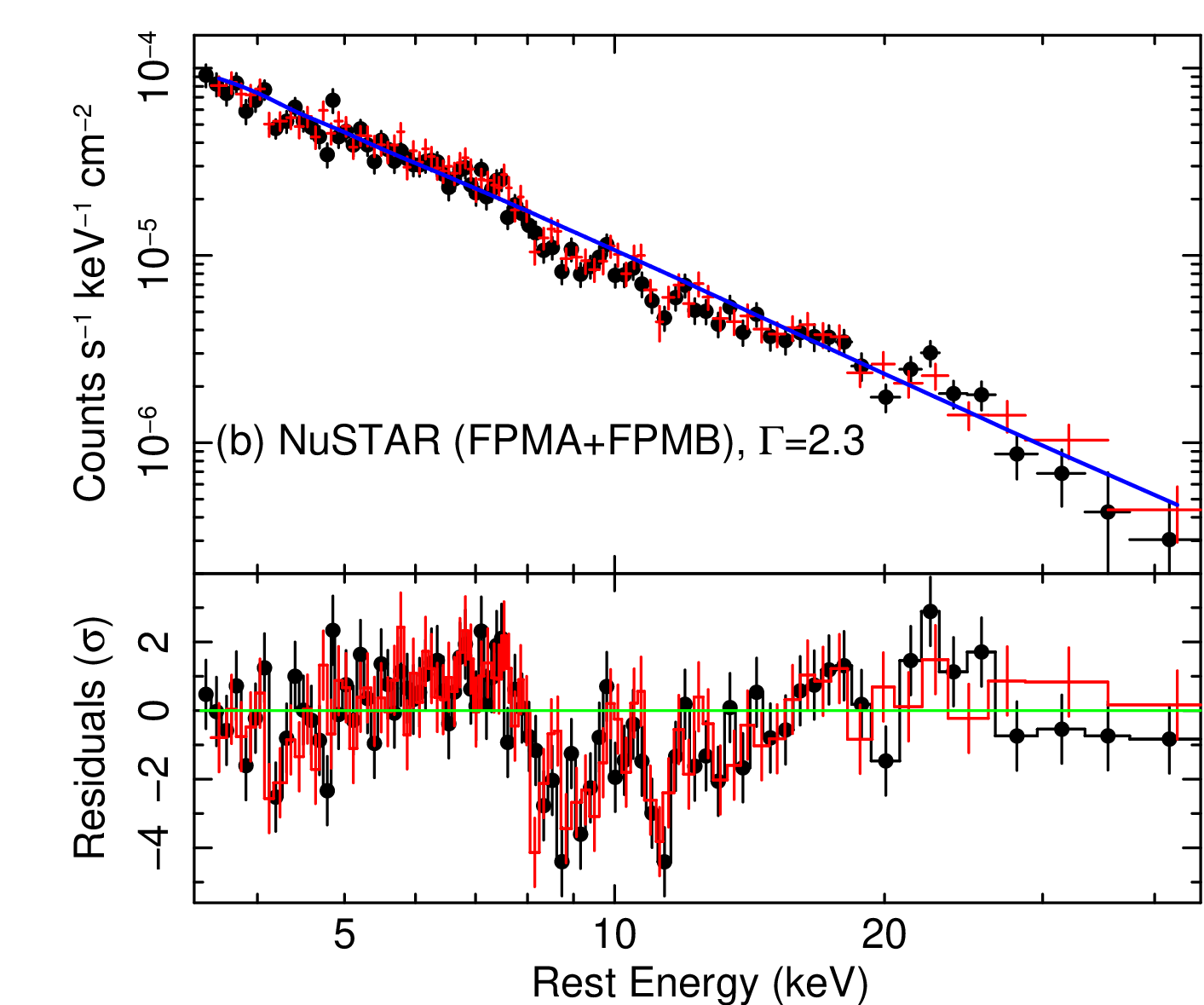}}
\end{center}
\caption{{\it NuSTAR} observations of PDS\,456 in March 2017. Panel (a) shows the background subtracted lightcurve of PDS\,456 over the 3--40\,keV band. The duration of the two simultaneous {\it XMM-Newton} observations (OBS\,1, OBS\,2) are marked in red. Note OBS\,1 is coincident
with a pronounced dip in the lightcurve.
Panel (b); {\it NuSTAR} FPMA (black circles) FPMB (red crosses) background subtracted spectra of PDS\,456, compared to a power-law (blue line) of $\Gamma=2.3$. The lower panel shows the residuals against the power-law. Significant absorption features 
are observed over the iron K band, at 8.9\,keV and 11.4\,keV (QSO rest frame).} 
\label{fig:nustar}
\end{figure}

PDS\,456 was subsequently observed with {\it NuSTAR} from March 23--26, 2017, with a total duration of 305\,ks. This coincided with two simultaneous {\it XMM-Newton} observations, hereafter OBS\,1 and OBS\,2, 
taken over two consecutive satellite orbits (see Table~1) and both in Large Window mode for EPIC-pn.
All data were processed using the \textsc{nustardas} v1.7.1, 
{\it XMM-Newton} \textsc{sas} v16.0 and \textsc{heasoft} v6.20 software.
{\it NuSTAR} source spectra were extracted using a 50\arcs\ circular region centered on the source and background from a 65\arcs\ circular region clear from stray light. 
{\it XMM-Newton} EPIC-pn spectra were extracted from single and double events, using a 30\arcs\ source region and $2\times34\arcs$\ background regions on the same chip. 
All spectra are binned to at least 50 counts per bin, while the background rates 
are $<10$\% of the net source rates. 
The 3-40\,keV {\it NuSTAR} lightcurve is shown in Figure~\ref{fig:nustar}, where OBS\,1 coincided with a pronounced dip in the source count rate $\sim50$\,ks into the {\it NuSTAR} observation.

Note that outflow velocities are given in the rest-frame of PDS\,456 at $z=0.184$, 
after correcting for relativistic Doppler shifts. Errors are quoted at 90\% confidence for one interesting parameter (or $\Delta\chi^2=2.7$).

\section{Spectral Analysis}

Initially we analyzed the time-averaged {\it NuSTAR} spectrum.
The count rate spectra from the FPMA+FPMB modules are shown 
in Figure~\ref{fig:nustar}, where the cross normalizations of FPMA and FPMB agree to within $\pm5$\% of each other. This is compared to a 
power-law of photon index of $\Gamma=2.3$ and
corrected for Galactic absorption, where $N_{\rm H}=2.4\times10^{21}$\,cm$^{-2}$, \citep{Kalberla05}.
PDS\,456 has a low flux throughout the 2017 observations, where the
mean 2-10\,keV flux of $2.6\times10^{-12}$\,erg\,cm$^{-2}$\,s$^{-1}$ is at the low end of the range 
measured in previous observations by \citet{Nardini15} and \citet{Matzeu17}.

While the hard X-ray continuum is well described by a power-law, the overall fit is very poor, with $\chi_{\nu}^{2}=327.1/169$, rejected with a null hypothesis probability of $P_{\rm N}=3.6\times10^{-12}$. Two strong absorption profiles are present in the residuals of both detectors between $8-12$\,keV. Modeling these with Gaussian profiles gave rest frame centroid energies of $E=8.93\pm0.15$\,keV 
and $E=11.4\pm0.3$\,keV, with equivalent widths of $EW=-430\pm80$\,eV and $EW=-380\pm100$\,eV 
respectively.
The addition of both lines significantly improved the fit by 
$\Delta\chi^2=-61.7$ and $\Delta\chi^2=-39.3$. 
The lower energy line is consistent with the energy of the Fe K absorption profile measured previously
in PDS\,456 \citep{Nardini15,Matzeu17}, however 
the second line appears at a substantially higher energy.
The lines are broadened and were fitted with a common velocity width, of $\sigma_{v}=20\,000^{+8\,000}_{-4\,000}$\,km\,s$^{-1}$, corresponding to $\sigma=600^{+250}_{-130}$\,eV at 8.93\,keV,  consistent with the width measured by \citet{Nardini15} previously.


If we associate the two lines with the Ly$\alpha$ and Ly$\beta$ transitions from H-like Fe at 6.97\,keV and 8.27\,keV respectively, then the inferred outflow velocities are inconsistent, with $v_{\rm out}=-0.24\pm0.02c$ and 
$v_{\rm out}=-0.31\pm0.02c$.  Note it would also be unusual for a higher order line to have such a high 
equivalent width as is observed here. 
Alternatively, the higher energy feature may be associated with an Fe K absorption edge. This gave a threshold energy of $E=10.7\pm0.2$\,keV, resulting in an acceptable fit. However, the velocity inferred from the edge is also inconsistent; e.g. for H-like Fe, a K-shell edge (at 9.27\,keV) blue-shifted to 10.7\,keV gives $v_{\rm out}=-0.14\pm0.02c$, while the Fe\,\textsc{xxvi} Ly$\alpha$ line (at 6.97\,keV) blue-shifted to 8.9\,keV gives $v_{\rm out}=-0.24\pm0.02c$. The velocities are also inconsistent if the two features instead arise from He-like iron. 

Given the lack of a plausible identification at a self consistent velocity, the two absorption lines are likely to arise from two outflowing systems with 
different velocities. If they are both associated with Fe\,\textsc{xxvi} Ly$\alpha$, then 
the outflow velocities are $v_{\rm out}=-0.24\pm0.02c$ and $v_{\rm out}=-0.46\pm0.03c$ (with 
slightly higher velocities inferred for He-like Fe). Thus while the lower energy line is consistent with the 
outflow velocities usually measured at iron K in PDS\,456 \citep{Nardini15,Matzeu17}, 
which are typically $0.25-0.3c$, the higher energy feature may originate from a new, 
very fast component of the wind.


\subsection{Photoionization Modelling} \label{sec:photoionization_modelling}

To test this, we fitted the {\it NuSTAR} spectrum with a self consistent \textsc{xstar} 
photoionization model, 
which accounts for any weak higher order lines and edges in addition to the 
strong $1s\rightarrow 2p$ resonance absorption. 
We adopt the same absorption model grids used in \citet{Nardini15}, 
where the optical to X-ray SED of PDS\,456 was used as the input continuum, which has 
an ionizing (1-1\,000\,Ryd) luminosity of $L_{\rm ion}=5\times10^{46}$\,erg\,s$^{-1}$. A turbulence velocity width of 15\,000\,km\,s$^{-1}$ was used, consistent with the Gaussian line width.
Solar abundances of \citet{GrevesseSauval98} were used throughout.
As the FPMA and FPMB spectra were consistent within errors, these were combined using \textsc{mathpha} into a single mean {\it NuSTAR} spectrum to maximize S/N. The response files were  combined using equal weightings for both modules, while the source to background area (\textsc{backscal}) scaling factors were propagated through to the combined spectrum.

\begin{figure}
\rotatebox{-90}{\includegraphics[width=9cm]{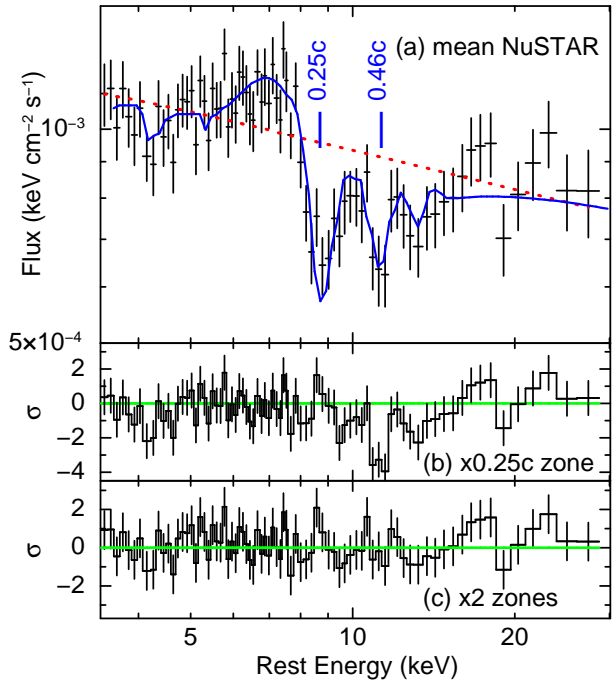}}
\caption{The mean 2017 {\it NuSTAR} spectrum of PDS\,456, fitted with a photoionization model (blue-line) consisting 
of two absorption zones with velocities, $-0.25c$ and $-0.46c$, which model the two absorption troughs at 9 and 11\,keV.  Note the third 
weak trough is due to a blend of higher order Fe K absorption. The underlying power-law 
continuum is plotted (red dotted line), while emission from the wind is observed between 6--8\,keV. 
Panel (b) shows the residuals against 
a model comprising of only the slower $-0.25c$ zone and the broad emission, while panel (c) is against the best-fit dual velocity 
zone model. The spectrum has been fluxed against a power-law  
and the best-fit model overlaid thereafter, it is not unfolded against the absorption model.}
\label{fig:mean_xstar}
\end{figure}

The best-fit \textsc{xstar} model fitted to the mean PDS\,456 spectrum is shown in Figure~\ref{fig:mean_xstar} (panel a).
Two outflowing absorption zones are required, one slower zone with $v_{\rm out}=-0.25\pm0.02c$ and the 
faster zone with $v_{\rm out}=-0.46\pm0.02c$; see Table\,1 for details. 
These velocities are consistent with the Gaussian analysis and the high ionization 
parameter of the absorbers ($\log\xi=5.5$) is consistent with the absorption 
lines arising from H-like iron (Fe \textsc{xxvi} Ly$\alpha$). 
Figure~\ref{fig:mean_xstar} (panel b) shows the residuals against a model including only the 
slower $-0.25c$ zone, which leaves significant residuals around 11\,keV and can then only be modeled by the 
fast $0.46c$ zone (see panel c).
Indeed, both absorption zones are required at $>99.99$\% 
confidence (with $\Delta\chi^2=-80.3$, slow zone and $\Delta\chi^2=-29.1$, fast zone), resulting in an acceptable fit statistic of $\chi_{\nu}^2=177.9/161$. 
As a final consistency check, we attempted to model the high energy feature with a lower ionization partial coverer, having the same velocity as the $-0.25c$ zone, but where the K-shell edge is blueshifted to higher energies. 
This single velocity absorber is rejected as the fit statistic is worse by $\Delta\chi^2=28.2$ (for $\Delta \nu=2$) compared to the two velocity model.

The spectrum shows significant excess emission, observed redwards 
of the absorption lines between $6-8$\,keV (see Figure~\ref{fig:mean_xstar}). 
This was first measured by \citet{Nardini15}, who resolved the broad P-Cygni profile at Fe K from 
PDS\,456. As per \citet{Nardini15}, the emission was modeled 
with an additive \textsc{xstar} emission grid and convolved with a Gaussian profile of 
width $\sigma=0.6^{+0.3}_{-0.2}$\,keV (or $\sigma_{\rm v}\sim24\,000$\,km\,s$^{-1}$).
The high equivalent width of the emission, with ${\rm EW}\sim350$\,eV, is consistent with the wide angle wind characterized by \citet{Nardini15}.

\subsection{The {\it XMM-Newton} Observations } \label{sec:emission_components}

The spectra obtained from the two {\it XMM-Newton} observations, OBS\,1 and OBS\,2, were also analyzed. 
Simultaneous {\it NuSTAR} spectra were extracted using identical good time intervals and then fitted jointly with 
their corresponding {\it XMM-Newton} spectra. The combined FPMA and FPMB spectra were used after first checking that the two modules were consistent. During the OBS\,1 spectrum, PDS\,456 was caught at an even lower flux ($F_{2-10\,keV}=1.8\times10^{-12}$\,erg\,cm$^{-2}$\,s$^{-1}$) coincident with the dip in the lightcurve. Note this may correspond to a short-lived absorption event, which will be discussed in a forthcoming paper on the broad band spectra.

\begin{figure}
\rotatebox{-90}{\includegraphics[width=9cm]{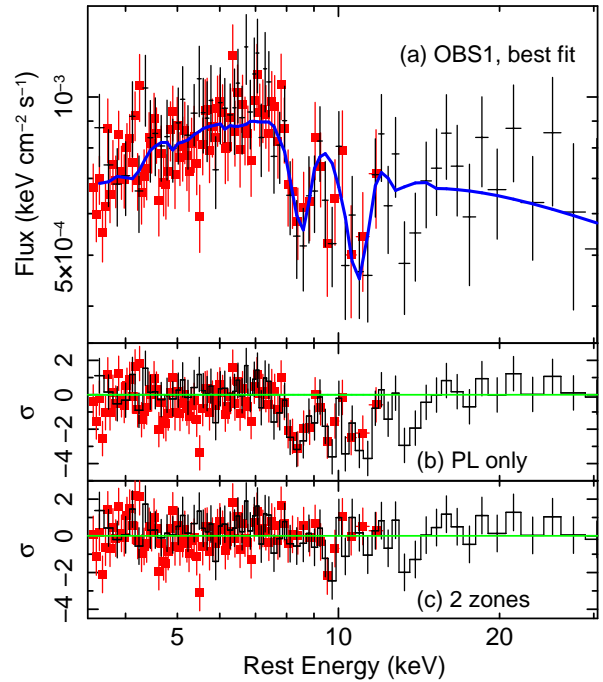}}
\caption{Simultaneous {\it XMM-Newton} EPIC-pn (red squares) and {\it NuSTAR} (black crosses) hard X-ray spectra, shown during the drop in flux during OBS\,1. Note the large depth of the high energy absorption line near 11\,keV. Panel (a) shows the fluxed spectra with the best fit dual absorber model overlaid. Panels (b) and (c) show the residuals against a power-law continuum only and the best-fit model respectively.}
\label{fig:obs1}
\end{figure}

The best-fit spectrum from OBS\,1 is shown in Figure~\ref{fig:obs1} and is well fitted by the dual velocity absorber.
Both absorption lines are detected at $>99.99$\% confidence, while the equivalent width of the high energy line is slightly stronger than the mean value, with ${\rm EW}=-530\pm150$\,eV. Note that the high energy feature is also independently detected in both the 
{\it NuSTAR} and {\it XMM-Newton} spectra, as the bandpass of the latter extends to 12\,keV in the QSO rest frame. 
For ease of comparison, we assumed a constant ionization 
across the mean, OBS\,1 and OBS\,2 spectra; see Table 1 for parameters.
The column density of the fast zone is slightly higher in OBS\,1 ($N_{\rm H}=5.0^{+1.4}_{-1.7}\times10^{23}$\,cm$^{-2}$), when compared to the mean ($N_{\rm H}=2.7^{+1.0}_{-1.2}\times10^{23}$\,cm$^{-2}$), 
to account for the increased depth of the high energy feature. 
The OBS\,2 spectrum is consistent with the mean and the faster zone appears somewhat weaker compared to the lowest flux OBS\,1 spectrum.

\section{Discussion} \label{sec:discussion}

Low flux observations of PDS\,456, obtained with {\it NuSTAR} and {\it XMM-Newton} in March 2017, 
have revealed a new relativistic component of the fast wind in this quasar, with $v=-0.46\pm0.02c$. 
Compared to studies of fast winds in other local, predominantly radio-quiet AGN \citep{Tombesi10,Gofford13}, this is the fastest wind detected to date in a nearby AGN. Perhaps the only other AGN with such an extreme fast wind is the high redshift ($z=3.9$) BAL QSO, APM\,08279+5255, where the outflow velocity may reach $\sim-0.7c$ \citep{Saez09, Chartas09, SaezChartas11}. Note, \citet{Hagino17} subsequently claimed the outflow velocity in APM\,08279+5255 may be somewhat lower, with $v\sim0.1-0.2c$. 

Here, the high velocity absorber appears considerably faster than all of the previous Fe K wind measurements in PDS\,456. \citet{Matzeu17} studied all twelve previous X-ray observations of PDS\,456, observed with either {\it XMM-Newton}, {\it NuSTAR} or {\it Suzaku}. They showed the wind velocity in PDS\,456 varies within the range $0.24c-0.34c$, while the strong positive correlation observed between the outflow velocity and X-ray luminosity was interpreted as possible evidence for a radiatively driven wind. 
The mean X-ray luminosity in 2017, of $L_{\rm 2-10 keV}=2.6\times10^{44}$\,erg\,s$^{-1}$, lies at the low luminosity end of the range observed by \citet{Matzeu17}, with $L_{\rm 2-10 keV}=2.8-10.5\times10^{44}$\,erg\,s$^{-1}$ and the velocity of the slower zone, with $v\sim-0.25c$ 
is consistent with the trend between velocity and luminosity. 

\begin{deluxetable}{lccc}
\tablecaption{The 2017 PDS 456 Observations and Outflow Parameters.}
\tablewidth{0pt}
\tablehead{
& \colhead{Mean} & \colhead{OBS\,1} & \colhead{OBS\,2}}
\startdata
Observations:-\\
Telescope & NuSTAR & XMM-Newton & XMM-Newton\\
Detector & FPMA+FPMB & EPIC-pn & EPIC-pn\\
OBSID & 60201020002 & 0780690201 & 0780690301\\
Start date & 2017/03/23 & 2017/03/23 & 2017/03/25\\
Start time & 05:31:09 & 19:25:01 & 06:27:09 \\
Exposure$^{a}$ & 157.0 & 39.6 & 64.9 \\  
Net rate (s$^{-1}$)$^{b}$ & $0.162\pm0.002$ & $0.774\pm0.004$ & $1.189\pm0.004$\\
\hline
Slower Zone:-\\
$N_{\rm H}$$^{c}$ & $4.2^{+1.3}_{-1.1}$ & $3.7^{+1.3}_{-1.1}$ & $3.9^{+1.0}_{-0.9}$\\
$\log\xi$$^{d}$ & $5.5^{t}$ & $5.5^{+0.3}_{-0.2}$ & $5.5^t$ \\
$v/c$  & $-0.25\pm0.02$ & $-0.21\pm0.02$ & $-0.27\pm0.02$ \\
$\Delta\chi^{2}$ & 80.3 & 24.0 & 42.2\\
\hline
Faster Zone:-\\
$N_{\rm H}$$^{c}$ & $2.7^{+1.0}_{-1.2}$ & $5.0^{+1.4}_{-1.7}$ & $2.9^{+1.4}_{-1.1}$\\
$\log\xi$$^{d}$ & $5.5^t$ & $5.5^t$ & $5.5^t$ \\
$v/c$  & $-0.46\pm0.02$ & $-0.43\pm0.02$ & $-0.43^t$ \\
$\Delta\chi^{2}$ & 29.1 & 23.6 & 10.4\\
\hline
Continuum:-\\
$\Gamma$ & $2.18^{+0.08}_{-0.05}$ & $2.41\pm0.09$ & $2.41^{t}$\\ 
$F_{\rm 2-10\,keV}$$^{e}$ & 2.58 & 1.84 & 2.54\\
\hline
$\chi_{\nu}^{2}$ & $177.9/161$ & $133.5/148$ & $246.5/255$
\enddata
\tablenotetext{a}{Net exposure after background screening and deadtime correction, in ks.}
\tablenotetext{b}{Net count rates, over 3--40\,keV for {\it NuSTAR} and 0.4--10\,keV for {\it XMM-Newton}.}
\tablenotetext{c}{Units of column density $\times10^{23}$\,cm$^{-2}$.}
\tablenotetext{d}{Ionization parameter (where $\xi=L/nR^{2}$) in units of erg\,cm\,s$^{-1}$.}
\tablenotetext{e}{Observed 2--10\,keV flux in units of $\times10^{-12}$\,erg\,cm$^{-2}$\,s$^{-1}$.}
\tablenotetext{t}{Denotes parameter is tied.}
\label{tab:xstar}
\end{deluxetable}

The fast, $\sim-0.45c$ component of the wind is not apparent 
in the past observations of PDS456, e.g. \citet{Nardini15} and see 
\citet{Matzeu17}, Figure~1. 
To confirm this, we checked all the previous {\it NuSTAR} observations of PDS\,456 for any significant higher energy absorption above 10\,keV, in addition to the persistent, but slower, $-0.25c$ wind component. However, aside from the 2017 observation, none was found. The most stringent constraint comes from the first {\it NuSTAR} observation of PDS\,456 in 2013 (hereafter, OBS\,A, \citealt{Nardini15}), where an upper-limit of $EW<80$\,eV is placed on any Gaussian absorption line profile above 10\,keV. Here the 2-10\,keV X-ray luminosity is $8\times10^{44}$\,erg\,s$^{-1}$, more than $\times3$ higher than in 2017. 
The upper limit on the column density of the fast zone during this high luminosity observation  
is $N_{\rm H}<1.1\times10^{23}$\,cm$^{-2}$ for a given ionization of $\log\xi=5.5$, significantly lower than for the low flux 2017 observations (see Table\,1).

The fast component may arise from an inner stream-line of an accretion disk wind \citep{ProgaKallman04,Sim10}, launched from very close to the black hole. Magneto-hydrodynamical mechanisms are also capable of accelerating winds up to these velocities, especially if the illuminating X-ray continuum is steep \citep{Fukumura10}, as is the case here with $\Gamma>2$. 
For a radiatively accelerated wind, the launch radius is:-
\begin{equation}
R_{\rm w} \approx 2 \left(\alpha \frac{L}{L_{\rm Edd}} - 1\right)\left(\frac{v_{\infty}}{c}\right)^{-2}
\end{equation}
where $R_{\rm w}$ is the wind launch radius in gravitational units ($R_{\rm G}$), $v_{\infty}$ is the terminal velocity and $\alpha$ is the force multiplier factor. 
In PDS\,456, with $L/L_{\rm Edd}=1$, $v_{\infty}=-0.45c$ and 
for a modest multiplier of $\alpha=2$, then $R_{\rm w}\sim10\,R_{\rm G}$ 
(or $\sim10^{15}$\,cm for a black hole mass of $10^{9}\,{\rm M}_{\odot}$). 
The maximum likely radial distance of the absorber can be derived under the assumption that $\Delta R / R <1 $, for a given wind streamline. For an ionizing 
luminosity of $L_{\rm ion}=5\times10^{46}$\,erg\,s$^{-1}$, $N_{\rm H}=5\times10^{23}$\,cm$^{-2}$ and 
$\log\xi=5.5$, then $R<L_{\rm ion}/N_{\rm H}\xi<10^{17}$\,cm
This is consistent with the radial distance of $R\sim10^{16}$\,cm estimated by \citet{Nardini15}, from the wind variability timescale in PDS\,456.

If the fast wind component is launched from close to the black hole and is fully exposed to the central X-ray source, then its ionization state will likely be very high. At these distances and for iron not to become fully ionized (where $\log\xi<6$), this requires densities of $n\sim10^{7} - 10^{11}$\,cm$^{-3}$ and which could be associated to matter lifted off the surface of the inner accretion disk. Given that this very fast wind component is detected during a low flux observation of PDS\,456, while it appears absent at higher fluxes (such as during OBS\,A in 2013), this may suggest that an inner wind zone would only be detected when its ionization state is low enough for the gas to not be fully ionized. 
Alternatively the innermost wind may be shielded by denser gas near the launch point, which could arise from the high column partial covering gas often seen in the lower flux X-ray spectra of PDS\,456 \citep{Matzeu16}.

Nonetheless, the fast zone may have important implications for the overall wind energetics.  For the $-0.25c$ wind in PDS\,456, \citet{Nardini15} estimated the kinetic power and mass outflow rate to be $\sim15$\% and $\sim50$\% of Eddington respectively.
As the power goes as $\dot{E}_{\rm kin}\propto \dot{m} v^{2} \propto v^{3}$, then the $-0.45c$ zone could require the wind power to be a factor of $\times6$ higher. Then the total kinetic power could reach Eddington, if the overall mass outflow rate of the fastest zone is similar to the slower one. In reality, it may be likely that we are observing a structured wind, with multiple velocity components, which are launched at different disk radii.  
Future high resolution X-ray calorimeter observations, with 
{\it XARM} and {\it Athena}, will be able to further reveal the velocity structure in high velocity winds such as PDS\,456.

\section{Acknowledgements}

JR acknowledges financial support through grants 
NNX17AC38G, NNX17AD56G and HST-GO-14477.001-A. AL acknowledges support via 
the STFC consolidated grant ST/K001000/1. EN is funded by the EU Horizon 2020 Marie Sklodowska-Curie 
grant no. 664931.
Based on observations obtained with XMM-Newton, an ESA science mission with instruments and contributions directly funded by ESA Member States and NASA.

\end{document}